\documentstyle[12pt,epsfig]{article}
\begin{document}
\begin{titlepage}
\rightline{August 4, 2000}

\bigskip
\begin{center}

\vfill
{\LARGE\bf The MONOLITH Project}

\vfill

\begin{figure}[h]
\begin{center}
\mbox{\epsfig{file=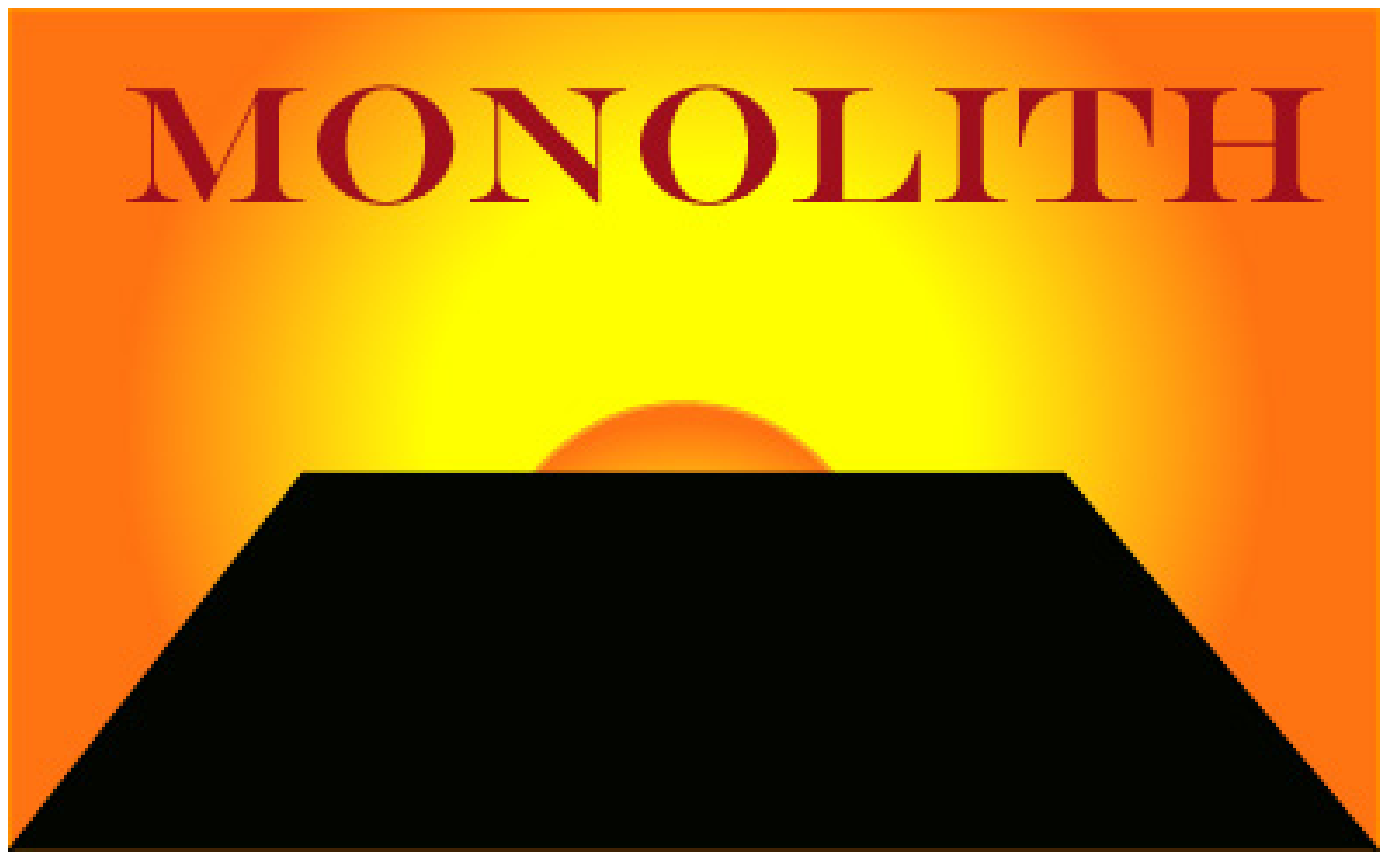,height=4cm}}
\end{center}
\end{figure}

\bigskip

{\large A. Geiser, Hamburg University}
\medskip

{\large for the MONOLITH Collaboration} 
\smallskip

({\bf M}assive {\bf O}bservatory for {\bf N}eutrino 
{\bf O}scillations or {\bf LI}mits on {\bf TH}eir existence)

\bigskip
\bigskip

{\large Bologna, Bonn, CNR Torino, Columbia, LNF Frascati, Hamburg, 
Humboldt Berlin, INR Moscow, L'Aquila, LNF Frascati, LNGS Gran Sasso, 
MEPhI Moscow, Milano, M\"unster, Napoli, Roma, Torino, Tunis}
     
\bigskip
\bigskip

\end{center}
\smallskip

\begin{abstract} 
\indent 
\textwidth 8.0cm 
\hoffset 5.0cm
MONOLITH is a proposed massive (34 kt) magnetized tracking calorimeter
at the Gran Sasso laboratory in Italy, optimized for the detection
of atmospheric muon neutrinos.
The main goal is to establish (or reject) the neutrino oscillation 
hypothesis through an explicit observation of the full first oscillation
swing. The $\Delta m^2$ sensitivity range for this measurement 
comfortably covers the 
complete Super-Kamiokande allowed region.
Other measurements include studies of matter effects and 
the NC/CC and $\bar{\nu}/\nu$ ratio, the study of cosmic ray muons
in the multi-TeV range, and auxiliary measurements from the 
CERN to Gran Sasso neutrino beam. 
Depending on approval, data taking with part of the detector could start in 
2004.
The detector and its performance are described, and its potential later use 
as a neutrino factory detector is addressed.
\end{abstract}

\vspace*{1cm}
{Contribution to the NuFACT'00 neutrino factory workshop,
  Monterey, California, USA, May 22-26, 2000} 

\end{titlepage}

\hoffset -0.7in 
\textwidth 6.0in 
\textheight 9.0in 
\normalsize 
\pagenumbering{arabic}

\begin{figure}[h]
\begin{center} \leavevmode \epsfig{file=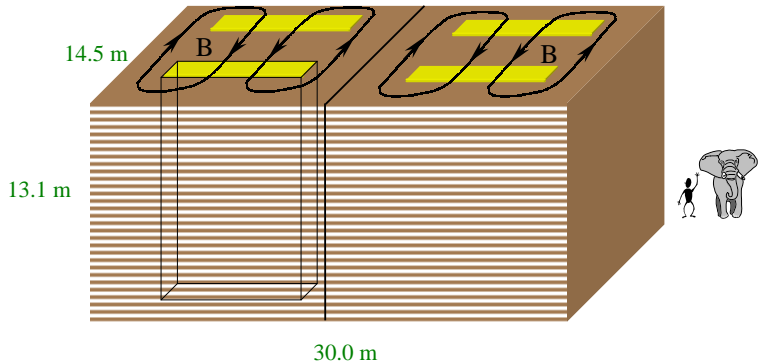,
   angle=-90, width=10cm}
\caption{\small Schematic view of the MONOLITH Detector.
The arrangement of the magnetic
field is  also shown.} \label{fig:module}
\end{center}
\vspace{-1cm}
\end{figure}

\section{Introduction}
\label{introduction}

While the cumulative evidence for neutrino oscillations is very
striking, the final proof that the observed anomalies are actually due
to neutrino  oscillations is still outstanding. In particular, the
current observations of atmospheric neutrinos \cite{SK-98,SK-2000}
are all consistent with
the hypothesis of maximal  $\nu_\mu$ oscillations, but do not yet
exclude some alternative unconventional  explanations
\cite{alternatives,decay}.
The main physics goal of the MONOLITH experiment \cite{monolith} 
is to establish the
occurrence of neutrino oscillations in atmospheric neutrinos through
the explicit  observation of the full first oscillation swing in
$\nu_\mu$ disappearance
\cite{Man98}, and
to investigate and presumably exclude alternative explanations.  This
also yields a  significantly improved measurement of the oscillation
parameters with respect to previous measurements.

The MONOLITH detector will be located at the Gran Sasso Laboratory
in Italy, and the measurement of the oscillation pattern can be
supplemented by measurements in the CERN to Gran Sasso neutrino beam.
A proposal is currently in preparation \cite{monolith}. If approved promptly, a
first part of the detector could be operational towards the end of 2004.  
The physics results described in the following sections correspond to an  
exposure of 4 years with the full detector.

\section{The MONOLITH detector}

The goals quoted above can be achieved with  a high-mass
tracking calorimeter with a coarse structure and magnetic field.
A large modular structure has been chosen for the detector (figure
\ref{fig:module}). One module  consists in a stack of 120
horizontal 8 cm thick iron planes with a surface area of $15\times
15\ {\rm m^2}$, interleaved with 2 cm planes of  sensitive
elements. The height of the detector is thus 12 meters. Thinner
plates, 2 and 4 cm thick, were also considered in the past,
however the 8 cm plate thickness resulted to be best compromise
between  physics result and detector costs. The magnetic field
configuration is also shown in figure \ref{fig:module}; iron
plates are magnetized at a magnetic induction of $\approx 1.3$ T.
The detector consists  of two modules.  Optionally, the downstream
module could be complemented by an end cap of  vertical planes to
improve the performance for non-contained muons from the CNGS beam. 
The total mass of the detector exceeds
34 kt.  
Glass Spark Counters (resistive plate chambers with glass
electrodes) have been chosen as active detector elements.
They provide two coordinates with a
pitch of 3 cm, and a time resolution of 2 ns.
Finally, an external veto made of scintillation counters reduces the
background from cosmic ray muons. 

\section{Observation of neutrino oscillation pattern}
\label{sect:pattern}

In the two flavour approximation, the survival probability for
neutrino oscillations in vacuum can be expressed by the well known
formula $P(L/E) = 1-\sin^2(2\Theta)\ \sin^2(1.27\ \Delta m^2\ L/E)$
where $L$ is the distance travelled in km, $E$ is the neutrino energy in GeV,
$\Theta$ is the neutrino mixing angle, and $\Delta m^2$ is the difference of
the mass square eigenvalues expressed
in eV$^2$.

\begin{figure}[h]
\begin{center}
\vspace{-.5cm}
\begin{tabular}{cc}
\mbox{\epsfig{file=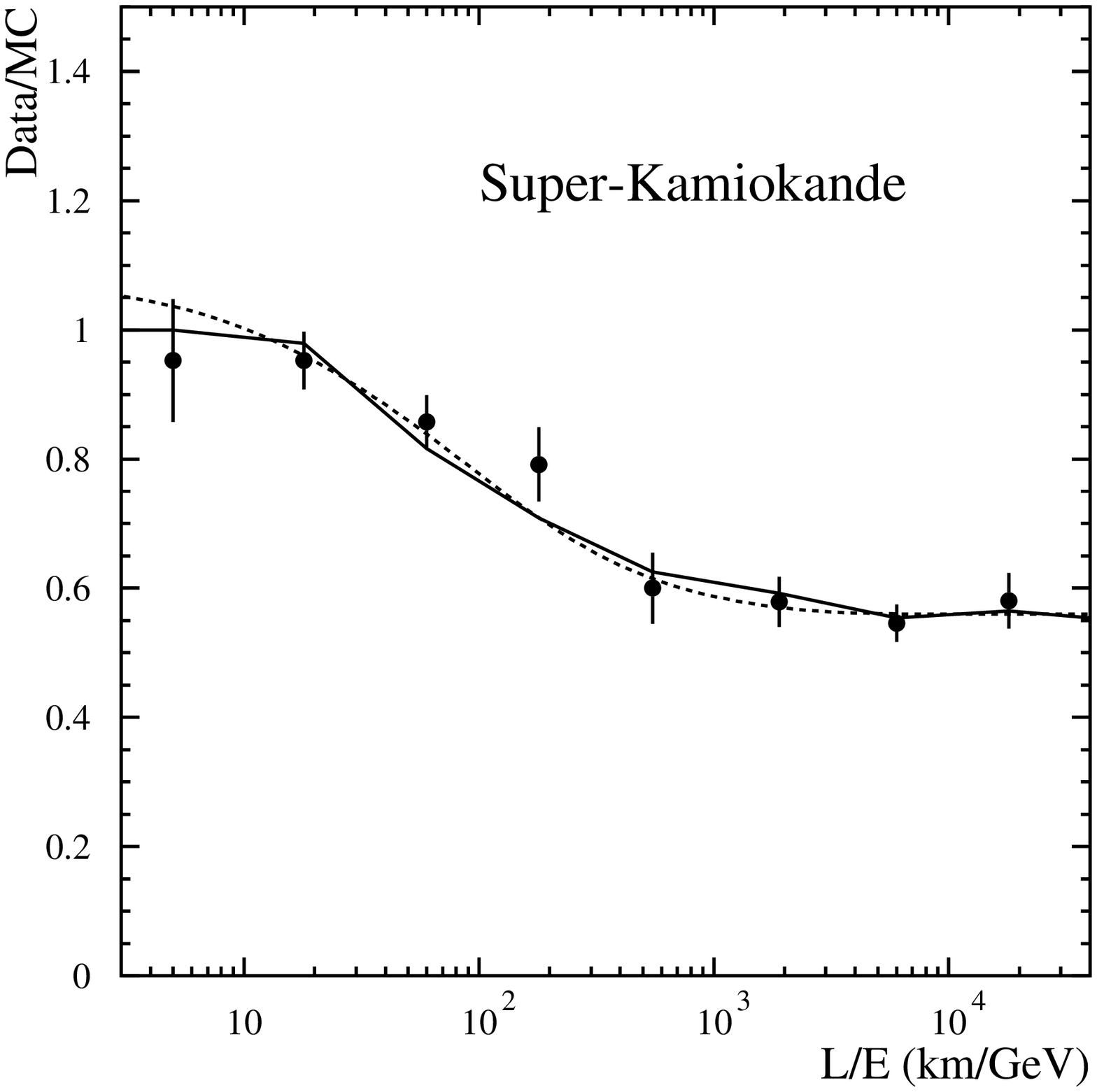,width=7cm}} &
\mbox{\epsfig{file=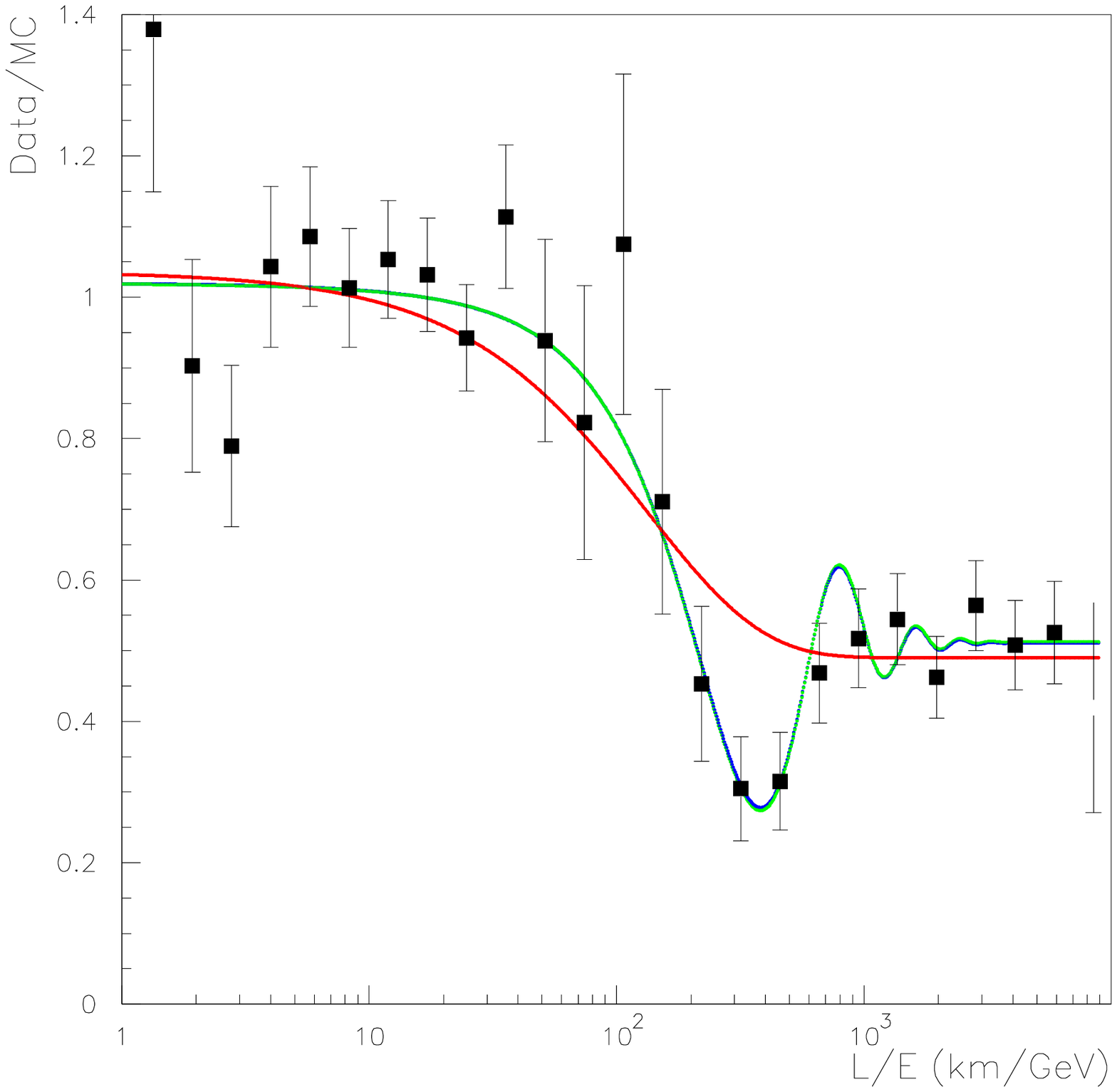,width=7cm}} \end{tabular}
\vspace{-.5cm}
\end{center}
\caption {\small {\it Left}: $L/E$ distribution from
Super-Kamiokande \cite{SK-2000} compared to the best fit oscillation
hypothesis (continous line), and to a
parametrization corresponding to the neutrino decay model of ref.
\cite{decay} (dashed line). The oscillations are smoothed out by detector
resolution.
{\it Right}:
$L/E$ distribution to be expected from MONOLITH for
$\Delta m^2 = 3 \times 10^{-3}$ eV$^2$ compared to the best fit
oscillation hypothesis (oscillating line) and to the corresponding
best fit of the neutrino decay model of ref. \cite{decay} (smooth threshold
effect).}
\label{fig:oscdip}
\end{figure}

However, none of the experiments which have yielded indications for neutrino
oscillations have so far succeeded to measure an actual sinusoidal oscillation
pattern. Figure \ref{fig:oscdip} shows the $L/E$ distribution of
Super-Kamiokande \cite{SK-2000} compared to the expectation for neutrino
oscillations and to a functional form suggested by
a recent neutrino decay model \cite{decay}. Once the detector resolution is
taken into account, the two hypotheses are essentially
indistinguishable \cite{decay}.
Even though the current evidence is very suggestive of neutrino oscillations,
a more precise measurement of the oscillation pattern
is the only way to actually prove the oscillation hypothesis for
atmospheric neutrinos.
The crucial issue here is to prove that muon neutrinos do not only
disappear,
but actually reappear at some larger L/E.

MONOLITH is explicitly designed to fill this
gap.  Having a similar mass (26 kt fiducial) as Super-Kamiokande, 
significantly larger
acceptance at high neutrino energies and better $L/E$ resolution, the
experiment is optimized to observe the full first oscillation swing,
including $\nu_\mu$ ``reappearance''.
Therefore, the oscillation hypothesis can be clearly distinguished
from other hypothesis which yield a pure disapperance threshold
behaviour (Figure \ref{fig:oscdip}).

\begin{figure}[h]
\begin{center}
\vspace{-.5cm}
\begin{tabular}{cc}
\mbox{\epsfig{file=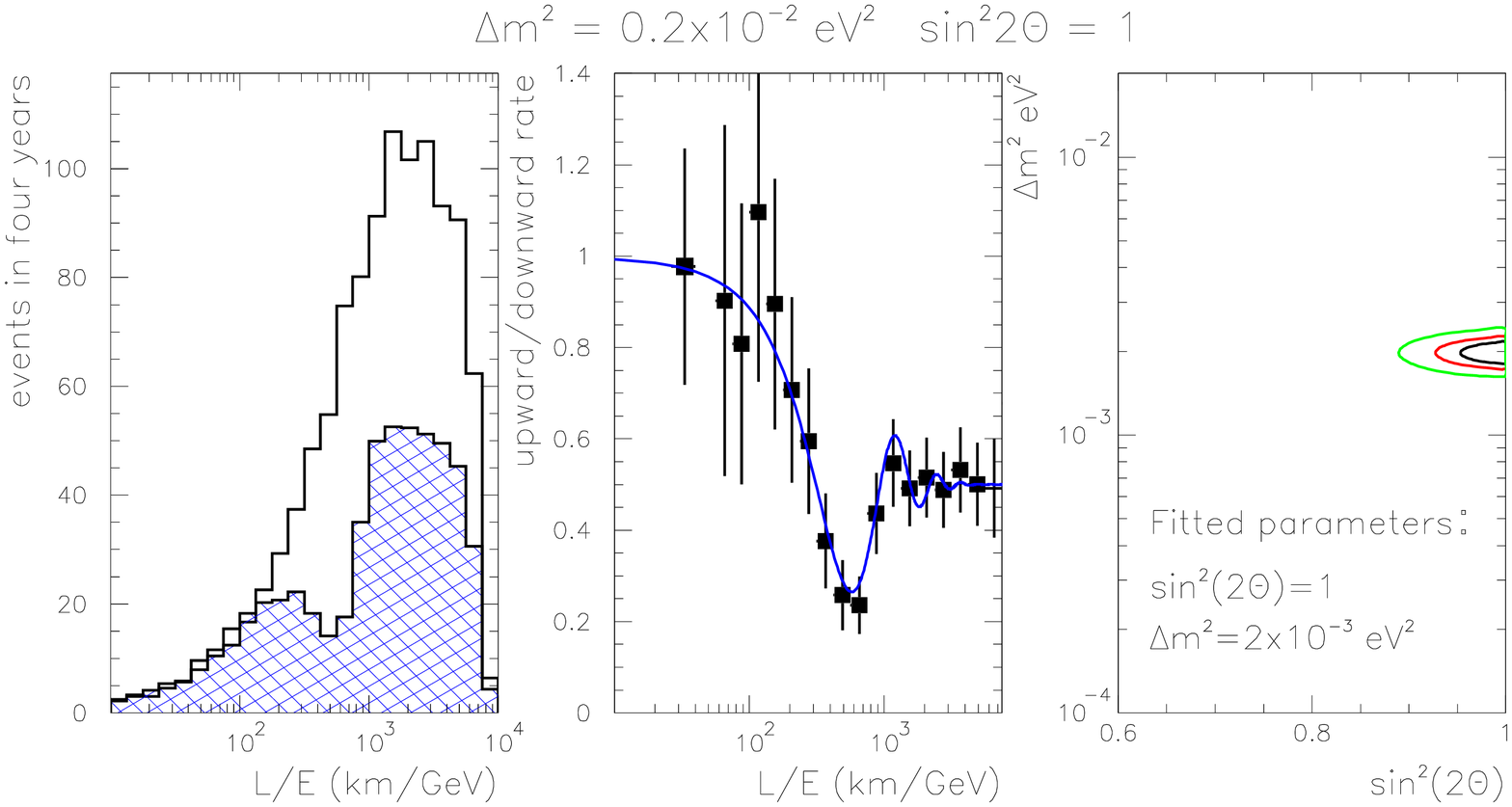,width=12cm}} &
\mbox{\epsfig{file=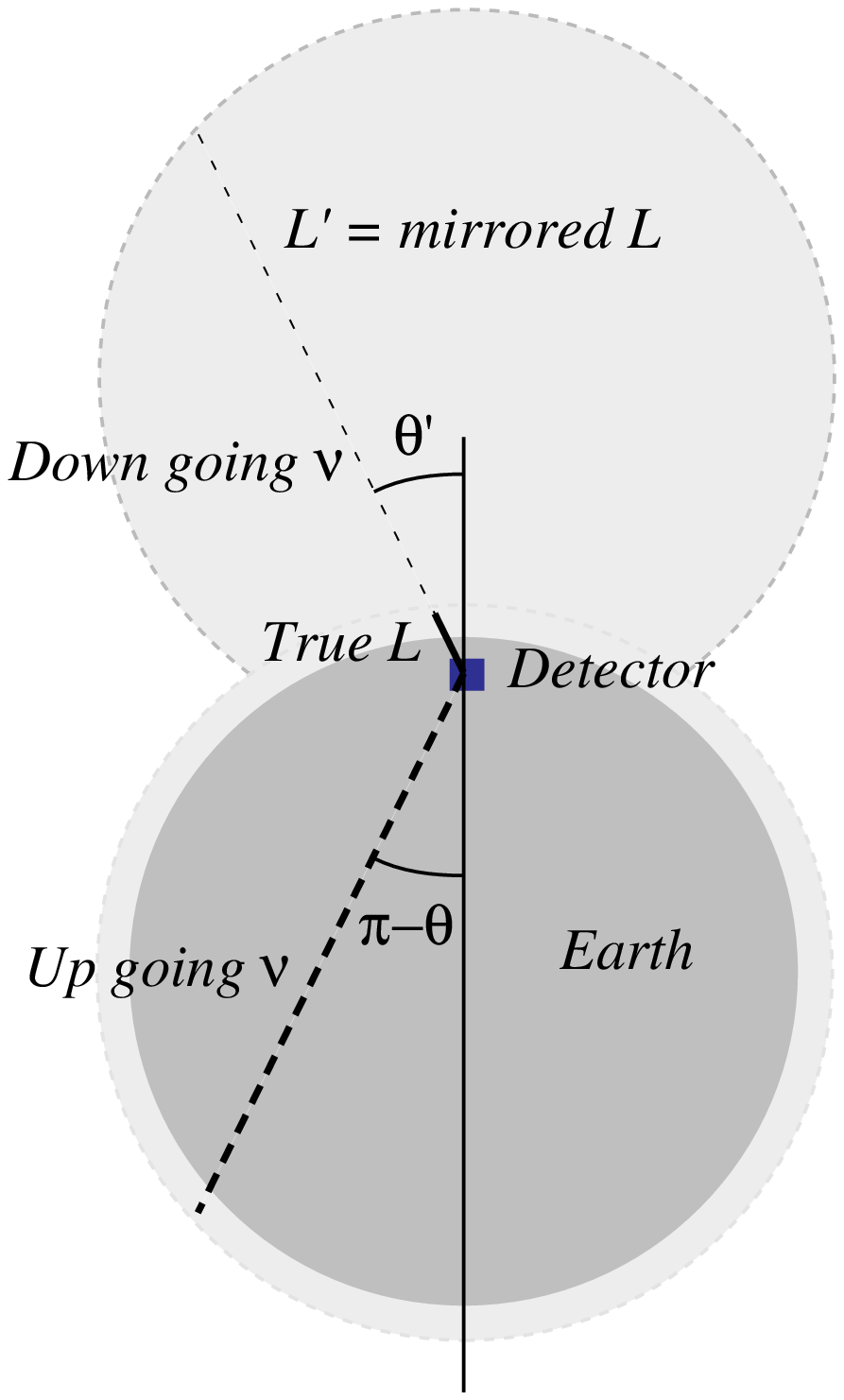,height=6cm}} \cr
\mbox{\epsfig{file=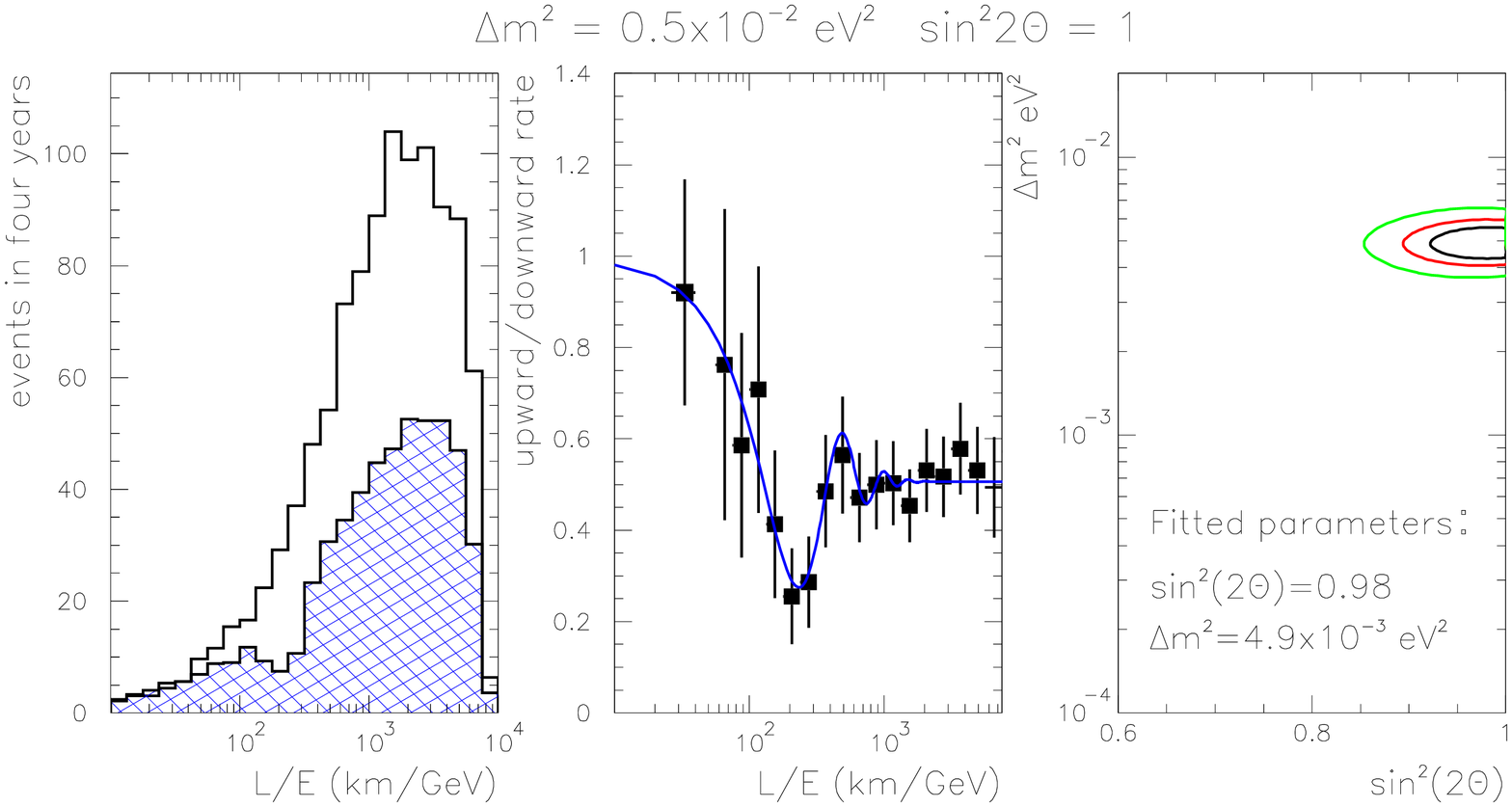,width=12cm}} & \cr
\end{tabular}
\vspace{-.5cm}
\end{center}
\caption{\small Results of the $L/E$ analysis on a simulated sample
  in the presence of
  $\nu_\mu\rightarrow\nu_x$   oscillations, with parameters $\Delta
  m^2 = 2\times 10^{-3}$ eV$^2$ and $\sin^2 (2\Theta)=1.0$ (top) and
  $\Delta m^2 = 5\times 10^{-3}$ eV$^2$ and $\sin^2(2 \Theta) = 1.0$.
  The figures show from left to right: %
  the $L/E$ spectrum of upward muon neutrino events (hatched area) and
  the $L/E$ ``mirrored'' spectrum of downward muon neutrino events
  (open area); their ratio with the best-fit superimposed 
  and the result of the
  fit with the corresponding allowed regions for oscillation
  parameters at 68\%, 90\% and 99\% C.L.;
artist's view of the mirror neutrino path length:
downward going neutrinos (zenith angle $\theta < \pi / 2$)
are assigned the distance they would have travelled if
$\theta = \pi - \theta$.}
\label{fig:one}
\end{figure}

\begin{figure}[t]
\begin{center}
\begin{tabular}{cc}
\mbox{\epsfig{file=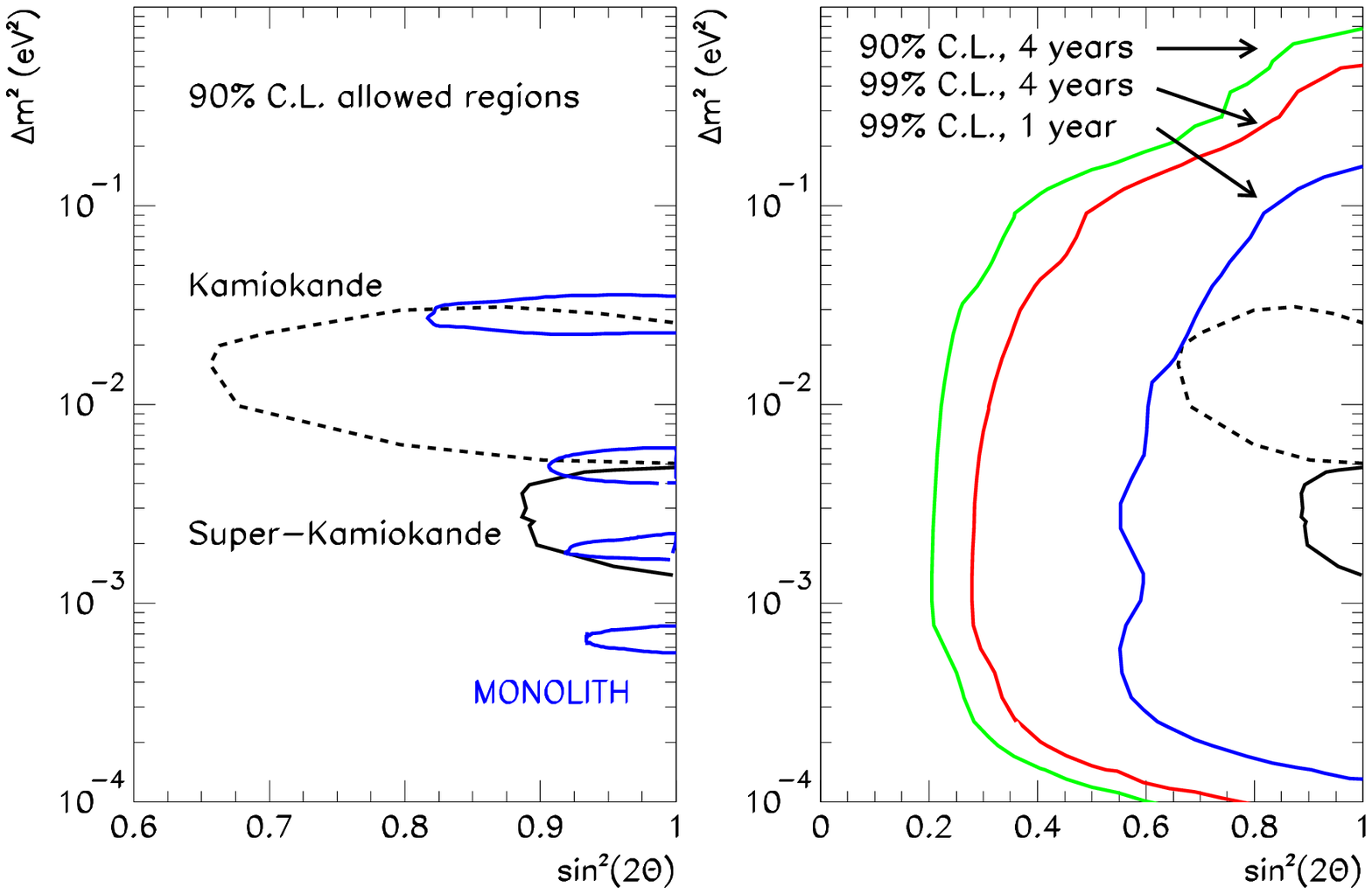,width=11cm}} &
\mbox{\hspace{-1cm}\epsfig{file=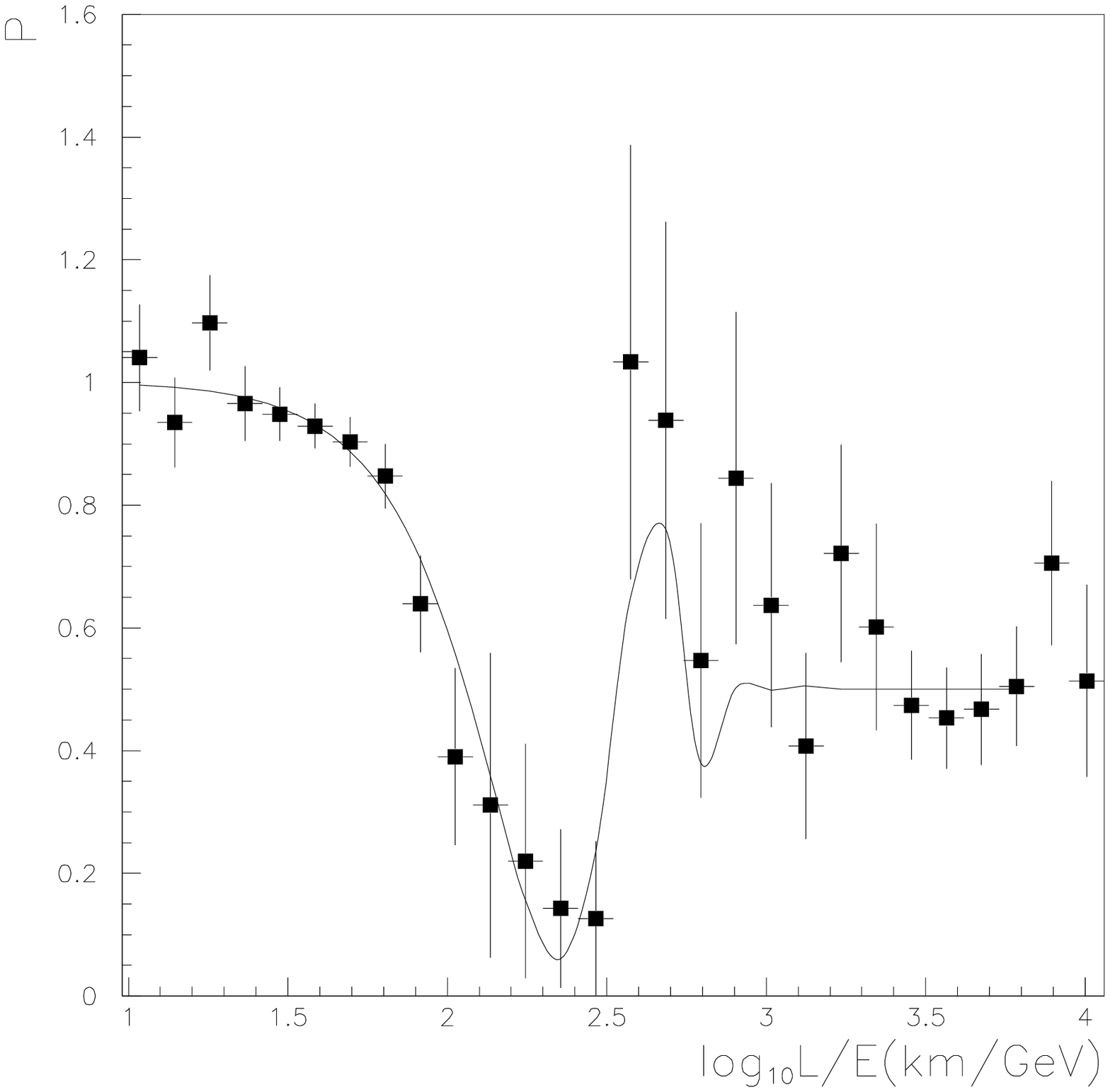,width=6cm} \vspace{1cm}}
\end{tabular}
\end{center}
\vspace{-.5cm}
\caption{\small {\it Left}: 
  Expected allowed regions of $\nu_\mu-\nu_\tau$ oscillation
  parameters for MONOLITH after four years of atmospheric neutrino exposure. 
  The results of the simulation  for $\Delta
  m^2 = 0.7, 2, 5, 30 \times 10^{-3}$ eV$^2$ and maximal mixing are
  shown. {\it Middle}: MONOLITH exclusion curves at 90\% and 99\% C.L. after
  one or 4 years of data taking assuming no oscillations. The full
  (dashed) black line shows the results of the
  Super-Kamiokande \cite{SK-2000}
  (Kamiokande \cite{Kamiokande}) experiment.
  {\it Right}: Example for the expected $L/E$ distribution ($\nu_\mu$
     survival probability) in MONOLITH
     with 2 years of data taking of atmospheric neutrinos and 1 year with
     the CERN-Gran Sasso neutrino beam for
     $\Delta m^2 = 5\times 10^{-3}$ eV$^2$ (points).
     Beam neutrinos dominate the L/E region
     below 10$^2$ km/GeV. For L/E $> 10^2$ km/GeV only
     atmospheric neutrinos contribute. Only statistical errors are shown.
}
\label{sensibilita}
\end{figure}

Furthermore, the sensitivity is almost independent of the oscillation
parameters (Fig. \ref{fig:one}).  This is in contrast to MINOS,
which can do a similar measurement at the highest allowed $\Delta m^2$
if the low energy beam is used
\cite{MINOSmin}, but has a hard time to observe a reappearance signal in
the lower $\Delta m^2$ range.
The good L/E resolution can be used to significantly
improve the measurement of the oscillation parameters over the full
allowed range (Fig. \ref{sensibilita}).
The systematic error can be reduced by comparing the upward neutrino
rate with the corresponding downward rate ``mirrored'' in L/E
(Fig. \ref{fig:one}).
Finally, if $\delta m^2$ is high, measurements of $\nu_\mu$ disappearance
in the CERN to Gran Sasso beam could complement the atmospheric neutrino
measurements (Fig. \ref{sensibilita}) if the systematic error can be
suitably controlled.

\section{Other physics topics} 

Provided that the neutrino oscillation hypothesis is confirmed,
another  goal of the experiment is to further investigate the nature
of these  oscillations. Depending on the oscillation parameters,
oscillations into  active ($\nu_\tau$) or sterile ($\nu_s$) neutrinos
can be distinguished  through their different effects on the up/down
ratio of neutral current (NC)-like  events, and/or through the
presence or absence of matter effects yielding a distortion of the
observed oscillation pattern as a function of energy  and/or muon
charge.

Even in the absence of sterile neutrinos, matter effects are present
in the case of a small contribution from $\nu_\mu - \nu_e$
oscillations at the ``atmospheric'' $\Delta m^2$. The corresponding MSW
resonance might be observable \cite{futatm} as a localized $\nu_\mu$
rate suppression either in $\nu_\mu$ or in $\bar\nu_{\mu}$.

Due to its ability of in situ measurement of the energy of every muon in the
multi-TeV range, MONOLITH will also be a unique facility for pioneer
investigations of cosmic ray muons in the
unexplored 100 TeV energy region.
The results of these studies
will give information which is relevant for the
solution of the problem of the knee in the cosmic ray energy spectrum.

Other potential physics topics include studies of the primary atmospheric
neutrino flux, the search for astrophysical point sources, and a search
for a neutrino ``line'' from WIMP annihilation in the center of the earth.

\section{MONOLITH at a neutrino factory}

Neutrino beams from future muon storage rings \cite{factories}
(neutrino factories) will be essentially
pure beams of either $\nu_\mu + \bar\nu_e$ or $\bar\nu_\mu + \nu_e$.
The occurence of $\nu_e - \nu_\mu$ or $\nu_e -\nu_\tau$ oscillations
would therefore manifest itself via the appearance of wrong sign muons.
A massive mag\-ne\-tized iron detector like MONOLITH, with good
muon charge separation and momentum measurement, could therefore be
well suited \cite{JJ} for the observation of such oscillations.
As pointed out in \cite{Rujula,Barger} this kind of beam
will in particular offer the possibility to measure the
$\theta _{13}$ mixing angle, currently only constrained
by the Super-Kamiokande and CHOOZ results, and the sign of $\Delta m^2$ through
matter effects. Depending on which of the solar neutrino solutions is
correct it might also open the way for the study
of CP violation in the neutrino system.
Interestingly, the optimization of detectors for the neutrino factory,
focusing on wrong sign muon appearance measurements, has yielded a detector
\cite{JJ} whose basic parameters are very similar to those of
MONOLITH. This is true in particular
when the source
is far enough away to impinge at a sizeable angle from below (horizontal
geometry of MONOLITH). For instance, a beam from Fermilab (L=7300 km) would
impinge at
an angle of 35$^o$, and be almost aligned with the Gran Sasso
hall axis, and therefore perpendicular to the magnetic field axis.
The results obtained in the physics studies of ref. \cite{NuFactsign}
concerning the measurements of $\theta_{13}$, sign of $\Delta m^2$, and CP
violation therefore qualitatively apply to MONOLITH used as a neutrino
factory detector.
Of course the potential timescale of a neutrino factory is quite different
from the one of the current atmospheric neutrino program.
Nevertheless, it might be interesting to consider that such a facility might
become reality within the lifetime of the MONOLITH project, and that its
useful life might be extended accordingly.

\section{Conclusions}

MONOLITH is a 34 kt magnetized iron tracking calorimeter proposed for
atmospheric neutrino measurements at the Gran Sasso Laboratory in Italy.
Its main goal is the proof of the neutrino oscillation hypothesis through
the explicit observation of a sinusoidal oscillation pattern 
($\nu_\mu$ reappearance).
Other goals include auxiliary measurements in the CERN to Gran Sasso beam,
and the investigation of potential $\nu_\mu - \nu_e$ and $\nu_\mu - \nu_s$
contributions. In the long term, the detector could also be used in a 
potential neutrino factory beam.


\begin{thebibliography}{000}

\bibitem{monolith} MONOLITH Progress Report, LNGS-LOI 20/99, CERN/SPSC 99-24,
               August 1999; \\
                   MONOLITH Proposal, LNGS P26/2000, CERN/SPSC 2000-031,
               August 2000; \\
               M. Ambrosio et al., The MONOLITH Prototype, Proceedings of the
               Bari RPC workshop, October 1999,
               ftp://netview.ba.infn.it/rpc/proceedings/gustavino.ps,
               submitted to Nucl. Instr. and Meth.

\bibitem{SK-98}
               Super-Kamiokande Collaboration, Y. Fukuda et al.,
               Phys. Rev. Lett. {\bf 81} (1998) 1562; \\
               Super-Kamiokande Collaboration, Y. Fukuda et al.,
               Phys. Lett. {\bf B 436} (1998) 33; \\
               Super-Kamiokande Collaboration, Y. Fukuda et al.,
               Phys. Lett. {\bf B 433} (1998) 9.

\bibitem{SK-2000} H. Sobel (Super-Kamiokande Collaboration), 
              Proceedings of XIX International Conference on Neutrino
              Physics and Astrophysics (Neutrino 2000), 
              Sudbury, Canada, June 16-21, 2000. 

\bibitem{alternatives} 
            R. Barbieri et al., P. Creminelli, and A. Strumia,
            IFUP-TH-2000-00, hep-ph/0002199, February 2000; \\
            E. Lisi, A. Marrone, and D. Montanino, 
                      hep-ph/0002053, February 2000. 
\bibitem{decay} V. Barger et al., Phys. Lett. {\bf B 462} (1999) 109.

\bibitem{Man98} G. Mannocchi et al., CERN/OPEN-98-004.
\bibitem{NOSEX} A. Curioni et al., hep-ph/9805249;\\
                M.~Aglietta et al., LNGS-LOI 15/98, CERN/SPSC 98-28, 
                SPSC/M615, Oct. 1998.

%
\bibitem{MINOSmin} S. Wojcicki, Proceedings of Neutrino '98, Takayama, Japan,
                   June 4-9, 1998.

\bibitem{Kamiokande} Kamiokande Collaboration, Y. Fukuda et al.,
                      Phys. Lett. {\bf B 335} (1994) 237.

\bibitem{beam} K. Elsener (editor), CERN 98-02 and INFN/AE-98/05, May 1998; \\
               R. Bailey et al., CERN-SL-99-034-DI, June 1999.

\bibitem{futatm} A. Geiser, Future Atmospheric Neutrino Detectors, 
                  Proceedings of XIX International Conference on Neutrino
              Physics and Astrophysics (Neutrino 2000),
              Sudbury, Canada, June 16-21, 2000.

\bibitem{NuFactsign}
         C. Albright et al., FERMILAB-FN-692, May 2000. \\
         V. Barger et al., hep-ph/0003184, March 2000,
                           hep-ph/0004208, April 2000. \\
         A. Cervera et al., hep-ph/0002108, February 2000. \\
         A. Bueno, M. Campanelli, and A. Rubbia, hep-ph/0005007, May 2000. \\
         M. Freund, P. Huber, and M. Lindner, TUM-HEP-373/00, hep-ph/0004085,
         April 2000.

\bibitem{factories} D. Cline and D. Neuffer,
                    AIP Conf. Proc. {\bf 68} (1980) 846;
                    reproduced in AIP Conf. Proc. {\bf 352} (1996) 10;\\
           S. Geer, Phys. Rev. {\bf D 57} (1998) 6989;
                    Erratum-ibid. {D 59} (1999) 039903.
\bibitem{JJ} J.J. Gomez-Cadenas and A. Cervera-Villanueva,
             talks given at $\nu$-Fact '99, Lyon, 5-9 July 1999.
\bibitem{Rujula} A. De Rujula, M.B. Gavela and P. Hernandez,
                   Nucl. Phys. {\bf B 547} (1999) 21.
\bibitem{Barger} V. Barger, S. Geer and K. Whisnant, hep-ph/9906487.

\end{thebibliography}
\end{document}